\documentclass[onecolumn,amsmath,amssymb,superscriptaddress,nofootinbib,11pt,a4paper]{revtex4}

\pdfoutput=1
\usepackage[T1]{fontenc}
\usepackage{CJK}
\usepackage{xcolor}
\usepackage{amsfonts}
\usepackage{epstopdf}
\usepackage{wrapfig} 
\usepackage{subcaption}
\usepackage{graphicx}  
\usepackage{dcolumn}   
\usepackage{bm}
\usepackage{float}
\usepackage[T1]{fontenc}
\usepackage{CJK}
\usepackage{xcolor}
\usepackage{amsfonts}
\usepackage{epstopdf}
\usepackage{wrapfig} 
\usepackage{subcaption} 
\usepackage{graphicx}  
\usepackage{dcolumn}   
\usepackage{bm}
\usepackage{float}
\usepackage[utf8]{inputenc}
\usepackage[T1]{fontenc}

\begin{document}

\title{Energy extraction from the accelerating Kerr black hole via magnetic reconnection in the plunging region and  circular orbit region}

\author{Ke Wang}
\email{kkwwang2025@163.com}
\affiliation{School of Material Science and Engineering, Chongqing Jiaotong University, \\Chongqing 400074, China}

\author{Xiao-Xiong Zeng\footnote{Electronic address: xxzengphysics@163.com.  (Corresponding author)}}
\affiliation{College of Physics and Electronic Engineering, Chongqing Normal University, \\Chongqing 401331, China}

\begin{abstract}
{Based on the magnetic reconnection mechanism, this study investigates how to extract energy effectively  from an accelerating Kerr black hole in the plunging region and circular orbit region. After introducing the properties of accelerating black holes, including the event horizon, ergosphere, circular orbits, and innermost stable circular orbit, we investigate the magnetic reconnection process in the plunging region. Specifically, we analyze variations of the  azimuthal angle with respect to the acceleration, examine changes in energy  per enthalpy of decelerated plasma, and plot energy extraction efficiency  along with permissible energy extraction regions. Results show that in the plunging region, at larger radii of reconnection locations, the accelerating black hole exhibits higher energy extraction efficiency than a Kerr black hole. Away from extremality, the acceleration parameter impedes energy extraction, while near extremality, it enhances extraction. We also study energy extraction in circular orbit region by plotting energy extraction efficiency  within permissible regions. We find that  the permissible energy extraction area is reduced and the efficiency exceeds that of Kerr black holes due to the existence of acceleration parameter. Larger acceleration parameters yield more effective energy extraction regardless of extremality, which is different from that in the plunging region. Additionally, energy extraction efficiency in the plunging region surpasses that in the circular orbit region, aligning with prior conclusions.
}
\end{abstract}

\maketitle
 \newpage
\section{Introdution}
Black holes are celestial objects predicted by Einstein's general relativity and play a crucial role in astrophysics. Recent astronomical observations have confirmed their existence \cite{LIGOScientific:2016aoc,EventHorizonTelescope:2019dse}. According to general relativity, rotating black holes contain significant amounts of extractable energy, making energy extraction from such black holes a longstanding topic of interest among physicists. The first proposed energy extraction process is known as the Penrose process \cite{Penrose:1969pc}. In this mechanism, negative energy orbits exist within the ergosphere. A particle with energy $E$ falls freely from infinity toward the black hole and splits into two fragments within the ergosphere. To an observer at infinity, one fragment carries negative energy and ultimately crosses the event horizon, while the other escapes along a geodesic with positive energy, thereby enabling energy extraction. However, the Penrose process is considered highly idealized \cite{Wald:1974kya}. Building upon Penrose's pioneering work, scientists have subsequently developed numerous alternative energy extraction mechanisms \cite{Teukolsky:1974yv,1975,Blandford:1977ds,1990}.

Based on astronomical observations, magnetic fields exist around black holes \cite{EventHorizonTelescope:2021bee,EventHorizonTelescope:2021srq,EventHorizonTelescope:2024hpu}. The renowned Blandford-Znajek mechanism utilizes these magnetic fields. Consequently, leveraging magnetic reconnection for energy extraction has gained significant popularity in recent years. Building upon early exploratory studies \cite{Koide:2008xr}, Comisso and Asenjo applied the magnetic reconnection approach to extract energy from Kerr black holes \cite{Comisso:2020ykg}. In rapidly rotating black holes, antiparallel magnetic field lines exist on the poloidal plane perpendicular to the equatorial plane \cite{14,15,16,17,18,19}. During the magnetic reconnection process, due to the change in the direction of the magnetic field on the poloidal plane near the equatorial plane, a current sheet emerges. When a current sheet exceeds a critical aspect ratio on this plane, plasma instabilities disrupt it \cite{20,21,22}. Subsequent plasmoid formation triggers rapid magnetic reconnection \cite{23,24,25}, converting magnetic energy into plasma kinetic energy and enabling plasma ejection from the reconnection layer. In this process, plasma in the current sheet separates into two components: one co-rotating with the black hole while accelerating, and the other counter-rotating while decelerating. Analogous to the Penrose process, to a distant observer the decelerated plasma carries negative energy and ultimately crosses the event horizon, whereas the accelerated component escapes to infinity with energy exceeding the initial input. Energy extraction is thereby achieved. As plasma moves beyond the reconnection zone, magnetic tension relaxes. Frame-dragging effects then restretch the magnetic field lines, reactivating magnetic reconnection. This establishes magnetic reconnection as an ongoing cyclic process. The Comisso-Asenjo process can yield higher power output than the Blandford-Znajek mechanism under certain conditions. Consequently, this approach has been rapidly extended to other rotating black holes \cite{26,27,28,29,30,31,32,33,34,35,36,37,38} and wormholes \cite{39}. Current research on magnetic reconnection energy extraction primarily focuses on circular orbit regions. Recently, Reference \cite{Chen:2024ggq} proposed energy extraction in the plunging region, where plasma initially follows circular orbits outside the innermost stable circle orbit (ISCO) but transitions to plunging motion from the ISCO inward due to orbital instability. Studies including \cite{Chen:2024ggq,41,42,43,44} have examined energy extraction in this plunging region. Reference \cite{41} first established under certain cases that plunging region energy extraction achieves higher efficiency than circular orbit regions. Therefore, investigating magnetic reconnection processes in the plunging region is both essential and necessary.

Within the framework of general relativity, black hole solutions describing strong gravitational fields and extreme spacetime structures hold central importance. Among these, the C-metric and its generalizations constitute a particularly significant family of solutions with profound physical implications and theoretical value \cite{45,46,47}. Originally discovered to describe black holes with intrinsic acceleration, this metric represents a key subclass or simplification of the Plebański-Demiański spacetime, which constitutes the most comprehensive family of exact solutions incorporating mass, angular momentum, electric charge, Newman-Unti-Tamburino (NUT) parameter, cosmological constant, and acceleration parameter \cite{48,49}. The physical interpretation of accelerating black holes is both intuitive and profound: they describe one or more black holes subjected to continuous external forces, commonly interpreted as tensions from cosmic strings or domain walls, beyond their own gravitational fields. This generates a nonzero intrinsic acceleration. Such external forces break the asymptotic flatness or (anti-) de Sitter symmetry of spacetime, introducing either conical deficits (cosmic strings) or acceleration horizons at infinity. Consequently, the black hole is no longer freely floating but undergoes acceleration along a specific direction. The introduction of acceleration fundamentally alters the causal structure and asymptotic behavior of black hole spacetimes. Properties of accelerating black holes have been extensively studied \cite{50,51,52,53,54} besides the magnetic reconnection. In this work, we investigate energy extraction via magnetic reconnection in both plunging and circular orbit regions of an accelerating Kerr black hole. In the plunging region, we find that at larger radii of reconnection locations, accelerating black holes exhibit higher energy extraction efficiency than Kerr black holes. Away from extremality, the presence of an acceleration parameter impedes energy extraction, while near extremality it enhances extraction. In  the circular orbit region, our results show that  the acceleration parameter reduces the permissible energy extraction area, and the efficiency  surpasses that of Kerr black holes. Additionally, we confirm that energy extraction efficiency in the plunging region exceeds that in circular orbits, consistent with established conclusions.

The remainder of this paper is organized as follows. Section 2 provides a brief description of accelerating Kerr black holes. Subsection 3.A outlines the magnetic reconnection process in the plunging region. Subsection 3.B analyzes energy extraction efficiency, while Subsection 3.C presents the permissible energy extraction regions. Subsection 4.A describes the magnetic reconnection mechanism in circular orbits. Subsection 4.B examines energy extraction efficiency in circular orbits, and Subsection 4.C maps the corresponding permissible extraction regions. We conclude our findings in Section 5.

\section{Introduction to accelerating Kerr Spacetime}
In this section, we briefly review accelerating Kerr black holes and introduce their key properties. In Boyer-Lindquist (BL) coordinates, the line element of an accelerating Kerr black hole can be expressed as \cite{Sui:2023rfh,Anabalon:2018qfv}
\begin{equation}
ds^2 = \frac{1}{H^2} \left\{ -\frac{\Delta_r}{\Sigma} \left( \frac{dt}{\alpha} - a \sin^2 \theta \, d\phi \right)^2 + \frac{\Sigma}{\Delta_r} dr^2 + \frac{\Sigma}{\Delta_\theta} d\theta^2 + \frac{\Delta_\theta}{\Sigma} \left( \frac{a \, dt}{\alpha} - (r^2 + a^2) d\phi \right)^2 \right\},
\end{equation}
where
\begin{equation}
\begin{gathered}
H = 1 + Ar \cos \theta,  \Sigma = r^2 + a^2 \cos^2 \theta, \Delta_r = (1 - A^2 r^2)(r^2 - 2Mr + a^2 ),\\
\Delta_\theta = 1 + 2MA \cos \theta + A^2a^2\cos^2 \theta,\alpha = \sqrt{\frac{1 - a^2 A^2}{1 + a^2 A^2}}.
\end{gathered}
\end{equation}
Here, $M$ represents the black hole mass, $a$ denotes the spin parameter, and $A$ is the acceleration parameter. When $A=0$, the metric reduces to the Kerr solution. Since this metric includes a conformal factor $H$ with dependencies on both $r$ and $\theta$, the geodesic equations for massive particles cannot be separated through the Hamilton-Jacobi formalism. Nevertheless, for the simplified magnetic reconnection processes typically studied on the equatorial plane ($\theta=\pi/2$) \cite{Comisso:2020ykg}, the radial equation for massive particles can be solved using the normalization condition. We adopt the framework from reference \cite{Comisso:2020ykg}, where reconnection is modeled in the equatorial current sheet. The normalization condition yields
\begin{equation}
g_{tt} \dot{t}^2 + 2g_{t\phi} \dot{t} \dot{\phi} + g_{rr} \dot{r}^2 + g_{\phi\phi} \dot{\phi}^2 = -1 ,  \label{3}
\end{equation}
Here, the dot  denotes differentiation with respect to proper time $\tau$. The energy $E$ and angular momentum $L$ are defined as
\begin{equation}
E = -g_{tt} \dot{t} - g_{t\phi} \dot{\phi}, L = g_{t\phi} \dot{t} + g_{\phi\phi} \dot{\phi} .\label{4}
\end{equation}
Solving these    equations yields
\begin{equation}
\dot{t} = \frac{\alpha^2 (E g_{\phi\phi} + L g_{t\phi})}{\Delta_r},
\dot{\phi} = -\frac{\alpha^2 (E g_{t\phi} + L g_{tt})}{\Delta_r} .   
\end{equation}
Substituting the normalization condition and metric components, we obtain
\begin{equation}
\dot{r}^2 = \frac{\Delta_r \left( -r^2 - \alpha^2 a^2 E^2 + 2 \alpha a E L - L^2 \right) + \left( \alpha E (r^2 + a^2) - a L \right)^2}{r^4}. \label{6}
\end{equation}
The horizon is determined by $\Delta_r=0$, yielding solutions
\begin{equation}
r_A = \dfrac{1}{A},r_+ = M + \sqrt{M^{2} - a^{2}},r_- = M - \sqrt{M^{2} - a^{2}} ,
\end{equation}
where $r_A$ denotes the acceleration horizon, $r_+$ the event horizon, and $r_-$ the Cauchy horizon. Generally, $r_- < r_+ < r_A$ holds. The conformal boundary occurs at $r_B=\frac{1}{|A\cos\theta|}$. For $\theta=\pi/2$, this reduces to $r_B=\infty$. The ergosphere boundary is defined by $g_{tt}=0$. On the equatorial plane, this condition takes the form
\begin{equation}
\left(1-A^2 r^2\right) \left(a^2-2 M r+r^2\right)=a^2.    
\end{equation}
This equation possesses four roots ordered as $r_{E-}<0<r_E<r_{EA}$. Here $r_{EA}$ represents the accelerating horizon ergosphere, whose value is slightly smaller than the acceleration horizon. In this work, we consider energy extraction from the event horizon ergosphere, specifically the third largest root of the equation. For massive particles on the equatorial plane, the effective potential satisfies
\begin{equation}
g_{rr} \dot{r}^{2} = V(r) .
\end{equation}
Expressed in terms of energy $E$ and angular momentum $L$ through equations \eqref{3} and \eqref{4}, the effective potential reads
\begin{equation}
 V(r) = \frac{g_{\phi\phi} E^{2} + 2g_{t\phi} E L + g_{tt} L^{2}}{g_{t\phi}^{2} - g_{tt} g_{\phi\phi}} - 1.   
\end{equation}
Circular orbits on the equatorial plane require \cite{
Zeng:2021dlj,Zeng:2021mok,Zeng:2022pvb}
\begin{equation}
  V(r) =0,V'(r)=0. \label{11}
\end{equation}
These conditions yield
\begin{equation}
\begin{aligned}
\Omega_{K} &= \frac{ -\partial_{r} g_{t\phi} \pm \sqrt{ (\partial_{r} g_{t\phi})^{2} - (\partial_{r} g_{tt})(\partial_{r} g_{\phi\phi}) } }{ \partial_{r} g_{\phi\phi} }, \\
E &= -\frac{ g_{tt} + g_{t\phi} \Omega_{K} }{ \sqrt{ -g_{tt} - 2g_{t\phi} \Omega_{K} - g_{\phi\phi} \Omega_{K}^{2} } }, \\
L &= \frac{ g_{t\phi} + g_{\phi\phi} \Omega_{K} }{ \sqrt{ -g_{tt} - 2g_{t\phi} \Omega_{K} - g_{\phi\phi} \Omega_{K}^{2} } }.  
\end{aligned}
\end{equation}
For the ISCO, condition \eqref{11} is supplemented by
\begin{equation}
 V''(r)=0.    
\end{equation}
Regarding energy extraction, only prograde orbits are considered. For retrograde orbits, the photon sphere (or ISCO) resides beyond the ergosphere, prohibiting energy extraction via circular motion. Concerning the plunging region, while particles can reach the ergosphere from the ISCO, the energy density per enthalpy at infinity remains positive for both accelerated and decelerated plasma flows. This positive value continues to preclude energy extraction \cite{Chen:2024ggq}. Throughout this paper, we adopt unit mass normalization with $M=1$.

\section{The Comisso-Asenjo process in the accelerating Kerr Spacetime in the plunging region}
\subsection{A Brief Description of the Comisso-Asenjo Process in the Plunging Region}
As shown in reference \cite{Chen:2024ggq}, using the perfect fluid approximation, the energy-momentum tensor can be written as
\begin{equation}
T^{\mu\nu}=\omega\,U^{\mu}U^{\nu}+p\,g^{\mu\nu}.    
\end{equation}
Here, $p$ and $\omega$ denote the fluid's proper pressure and enthalpy density respectively. For relativistic thermal plasmas, the relation $\omega = 4p$ holds. Meanwhile, $U^{\mu}$ represents the fluid's four-velocity. The metric can be expressed in the 3+1 formalism through the decomposition
\begin{equation}
 ds^{2}=g_{\mu\nu}dx^{\mu}dx^{\nu}=-\kappa^{2}dt^{2}+\sum_{i=1}^{3}\bigl {(}\sqrt{g_{ii}}dx^{i}-\kappa\beta^{i}dt\bigr{)}^{2}  , \label{15}
\end{equation}
where
\begin{equation}
\kappa=\sqrt{-g_{tt}+\frac{g_{t\phi}^{2}}{g_{\phi\phi}}}\,,\beta^{i}=\delta_{i\phi}\frac{\sqrt{g_{\phi\phi}}\,\omega^{\phi}} {\kappa},\omega^{\phi}=-g_{t\phi}/g_{\phi\phi}  .  
\end{equation}
Employing the Zero Angular Momentum Observer (ZAMO) frame \cite{Bardeen:1972fi}, the metric can be expressed through equation \eqref{15} as
\begin{equation}
ds^{2} = -d\hat{t}^{2} + \sum_{i=1}^{3} (d\hat{x}^{i})^{2}.   
\end{equation}
Consequently, the four-velocity of the fluid in the ZAMO frame relates to its BL counterpart through the transformation
\begin{equation}
\hat{U}^{\mu}=\hat{\gamma}_{s}(1,\hat{v}^{(r)} _{s},0,\hat{v}^{(\phi)}_{s})=(\frac{E-\omega^{\phi}L}{\kappa}, \sqrt{g_{rr}}\,U^{r},0,\,\frac{L}{\sqrt{g_{\phi\phi}}}),
\end{equation}
where
\begin{equation}
\hat{v}_{s}=\sqrt{\left(\hat{v}_{s}^{(r)}\right)^{2}+\left(\hat{v}_{s}^{(\phi)}\right)^{2}}  ,\hat{\gamma}_{s} = \frac{1}{\sqrt{1 - \hat{v}_{s}^{2}}}.
\end{equation}
Based on equation \eqref{6}, the radial component $U^{r}$ is expressed as
\begin{equation}
U^{r} =\dot{r}= -\sqrt{\frac{\Delta_r \left( -r^2 - \alpha^2 a^2 E_I^2 + 2 \alpha a E_I L_I - L_I^2 \right) + \left( \alpha E_I (r^2 + a^2) - a L_I \right)^2}{r^4}}. 
\end{equation}
Here $E_I$ and $L_I$ refer to values at the ISCO. In circular orbit regions, $U^{r}=0$, with only azimuthal velocity components present. Within the plunging region, the energy per enthalpy observed at infinity  for accelerated and decelerated plasmas can be expressed as
\cite{Chen:2024ggq,Comisso:2020ykg}
\begin{equation}
 \begin{aligned}  
\epsilon_{\pm} &=\kappa\hat{\gamma}_{s}\gamma_{out}\biggl{[}\bigl{(}1+\beta^{\phi }\hat{v}_{s}^{(\phi)}\bigr{)}\pm v_{out}\biggl{(}\hat{v}_{s}+\beta^{\phi}\frac {\hat{v}_{s}^{(\phi)}}{\hat{v}_{s}}\biggr{)}\cos\xi\mp v_{out}\beta^{\phi} \frac{\hat{v}_{s}^{(r)}}{\hat{\gamma}_{s}\hat{v}_{s}}\sin\xi\biggr{]}\\&
-\frac{\kappa}{4\hat{\gamma}_{s}\gamma_{out}\bigl{(}1\pm\hat{v}_{s}v_{out}\cos\xi\bigr{)}}\,, 
\end{aligned}\label{21}
\end{equation}
here, $v_{out}$ denotes the outflow plasma velocity measured in the fluid rest frame, and $\gamma_{out}$ represents the corresponding Lorentz factor. These quantities satisfy \cite{Chen:2024ggq}
\begin{equation}
v_{out}\simeq\sqrt{\frac{(1-{g}^{2})^{3}\sigma}{(1+{g}^{2})^{2}+(1-{g}^{2})^{3}\sigma}},{\gamma}_{out} = \frac{1}{\sqrt{1 - {v}_{out}^{2}}},
\end{equation}
where $\sigma$ denotes the upstream magnetization parameter and $g$ is the geometric index. In the high local magnetization limit with magnetic reconnection occurring at the maximum local rate, $g\simeq 0.49$ \cite{Chen:2024ggq}.
The azimuthal angle of magnetic field lines in the fluid rest frame, denoted by  $\xi$, satisfies \cite{Chen:2024ggq}
\begin{equation}
\xi=\arctan\left(\frac{\sqrt{g_{rr}g_{ \phi\phi}}\,\omega^{\phi}U^{r}}{\hat{\gamma}_{s}E_I-\kappa}\right).\label{23}
\end{equation}
This azimuthal angle $\xi$ is subject to the ideal megnetohydrodynamics(MHD) condition \cite{Ruffini:1975ne,Hou:2023bep}.

\begin{figure}[!h]
  \centering
  \begin{subfigure}{0.32\textwidth}
    \centering
    \includegraphics[width=\linewidth]{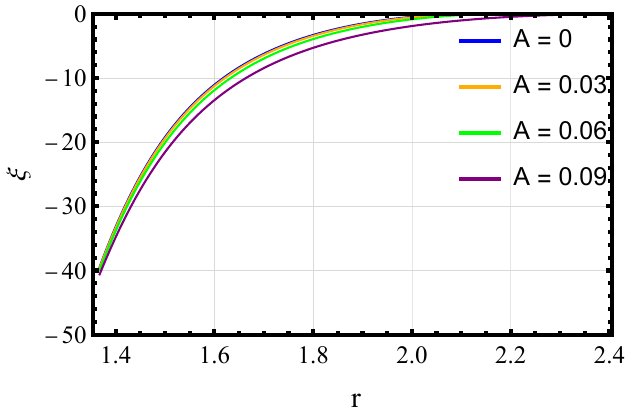}
    \caption{} 
  \end{subfigure}
  \begin{subfigure}{0.32\textwidth}
    \centering
    \includegraphics[width=\linewidth]{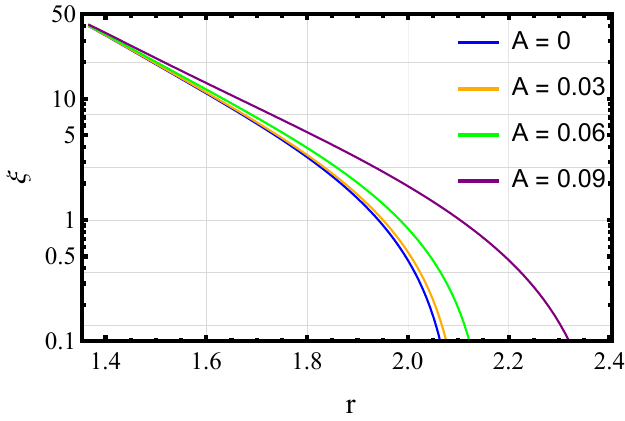} 
    \caption{} 
  \end{subfigure}
   \begin{subfigure}{0.32\textwidth}
    \centering
    \includegraphics[width=\linewidth]{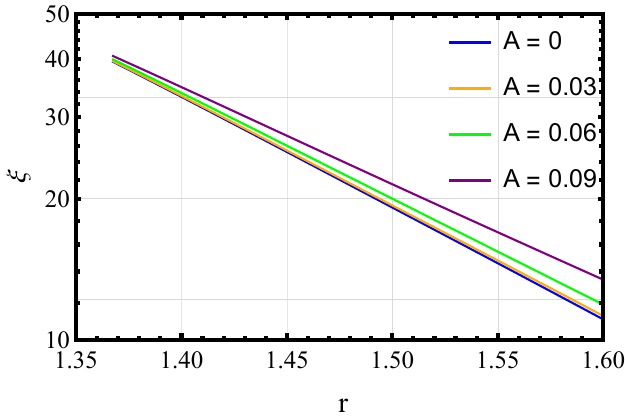} 
    \caption{} 
  \end{subfigure}
  \caption{(a) Variation of the azimuthal angle $\xi$ with $r$ for different values of $A$. (b) Logarithmic graph of the absolute value of $\xi$. (c) Logarithmic graph of the absolute value of $\xi$ when r is relatively low.}
  \label{fig:1}
\end{figure}

We plot the variation of the magnetic field azimuthal angle $\xi$ with  the radius of reconnection location $r$ for different values of $A$ in the left panel of Figure \ref{fig:1}, taking $a=0.93$. Additionally, to make the images clearer, we plotted the logarithm of the absolute value of $\xi$ in the middle panel of Figure \ref{fig:1}. However, since the logarithmic plot is less distinguishable at low $r$ values, we also plotted the logarithm of the absolute value of $\xi$ specifically for low $r$ values in the right panel of Figure \ref{fig:1}. As noted in \cite{Comisso:2020ykg}, the equatorial current sheet is susceptible to plasmoid instability \cite{20,21,22}, causing it to break up into numerous X-points. Among these, the dominant reconnection X-point is the one located at the intersection of the separatrices that encompass the global reconnection flow. This specific X-point governs the reconnection dynamics and defines the distance denoted as $r$.

\begin{figure}[!h]
\centering
\includegraphics[width=0.5\linewidth]{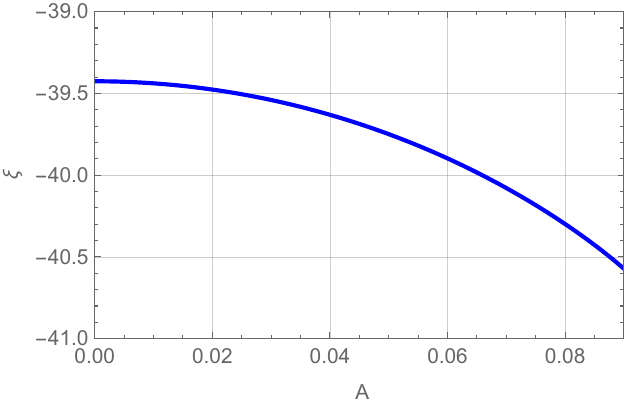}
\caption{Variation of the azimuthal angle  $\xi$  with $A$  at the event horizon. }
\label{fig:2}
\end{figure}

As can be seen from Figure \ref{fig:1}, $\xi$ is less than 0. At $r=$ISCO, $\xi$ equals 0. As $r$ decreases, that is, as we plunge inwards from the ISCO towards the event horizon, the absolute value of $\xi$ continuously increases. At the event horizon, the absolute value of $\xi$ reaches its maximum, which is consistent with the findings in reference \cite{Chen:2024ggq}. As $A$ increases, the absolute value of $\xi$ also continuously increases. We plot the value of $\xi$ at the event horizon as a function of $A$ in Figure \ref{fig:2}, which reflects this characteristic more clearly.

\begin{figure}[!h]
  \centering
  \begin{subfigure}{0.32\textwidth}
    \centering
    \includegraphics[width=\linewidth]{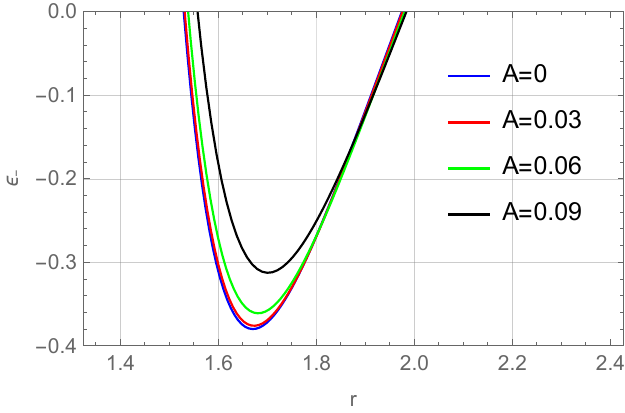}
    \caption{$a=0.93$} 
    \label{fig:3a}
  \end{subfigure}
  \begin{subfigure}{0.32\textwidth}
    \centering
    \includegraphics[width=\linewidth]{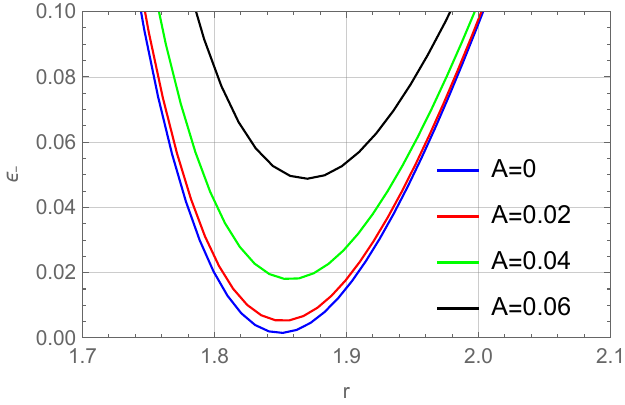} 
    \caption{$a=0.863$} 
    \label{fig:3b}
  \end{subfigure}
   \begin{subfigure}{0.32\textwidth}
    \centering
    \includegraphics[width=\linewidth]{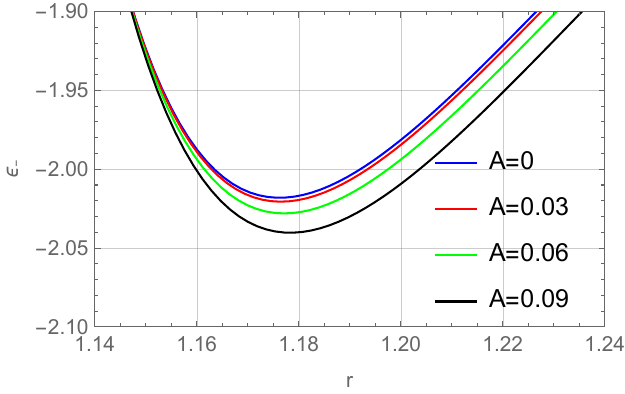} 
    \caption{$a=0.998$} 
    \label{fig:3c}
  \end{subfigure}
  \caption{Variation of $\epsilon_-$ with $r$ for different values of $A$. }
  \label{fig:3}
\end{figure}

For energy extraction to be possible, $\epsilon_-$ should be less than 0. We plot the variation of $\epsilon_-$ with $r$ for different $A$ values in Figure \ref{fig:3}. We take $\sigma=100$. The left panel is for $a=0.93$, the middle panel is for $a=0.863$ and the right panel is for $a=0.998$, corresponding to a near-extremal black hole case.
As can be seen from Figure \ref{fig:3}, $\epsilon_-$ decreases first and then increases with $r$. In cases not close to extremal black holes, such as $a=0.93$ and $a=0.863$, as the acceleration factor $A$ increases, the value of $\epsilon_-$ increases. This indicates that the presence of the acceleration factor hinders energy extraction. However, in the near-extremal black hole case, as the acceleration factor $A$ increases, the value of $\epsilon_-$ decreases. This suggests that for near-extremal black holes, the energy extraction capability of accelerated black holes surpasses that of Kerr black holes.

\subsection{Efficiency of Energy Extraction}

We  are interested  in  the efficiency of energy extraction in this subsection. The energy extraction efficiency is defined as \cite{Comisso:2020ykg}
\begin{equation}
\eta=\frac{\epsilon_{+}}{\epsilon_{+}+\epsilon_{-}}.\label{24}
\end{equation} 
For energy extraction to be meaningful, $\eta$ must exceed 1 since $\epsilon_{-}$ must be negative. We still take $\sigma=100$. The left panel corresponds to $a=0.93$, the middle panel to $a=0.998$, and the right panel shows a zoomed-in view of the middle panel.
\begin{figure}[!h]
  \centering
  \begin{subfigure}{0.32\textwidth}
    \centering
    \includegraphics[width=\linewidth]{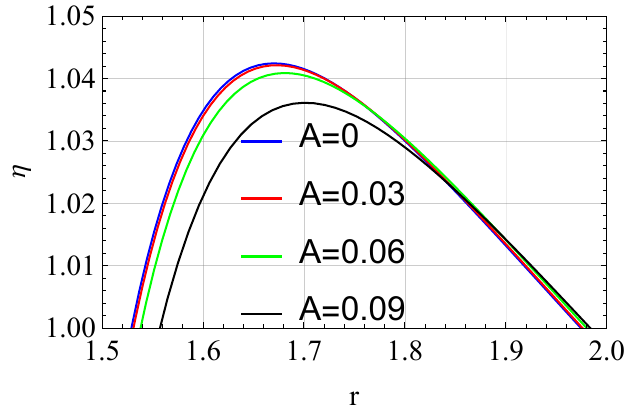}
    \caption{$a=0.93$} 
    \label{fig:4a}
  \end{subfigure}
  \begin{subfigure}{0.32\textwidth}
    \centering
    \includegraphics[width=\linewidth]{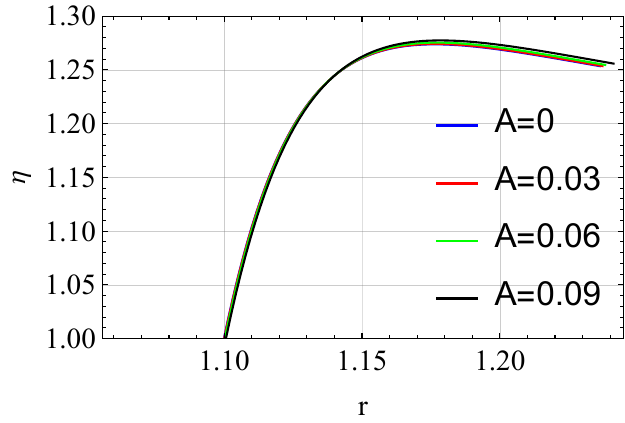} 
    \caption{$a=0.998$} 
    \label{fig:4b}
  \end{subfigure}
   \begin{subfigure}{0.32\textwidth}
    \centering
    \includegraphics[width=\linewidth]{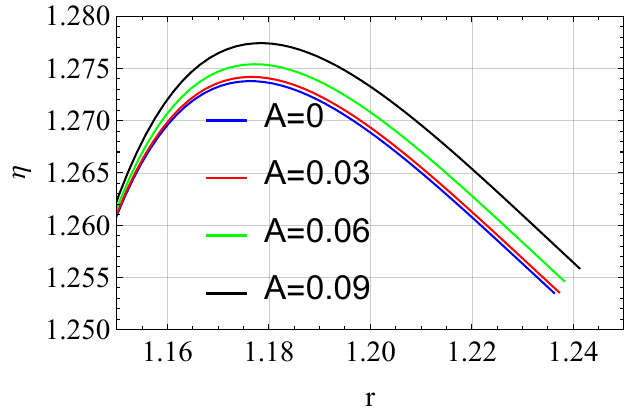} 
    \caption{Magnified view at $a=0.998$} 
    \label{fig:4c}
  \end{subfigure}
  \caption{Variation of energy extraction efficiency with $r$. }
  \label{fig:4}
\end{figure}
Figure \ref{fig:4} shows that away from extremality, efficiency exceeds the Kerr benchmark only at larger radii of reconnection locations when acceleration parameters are present. Typically, efficiency remains lower than Kerr black holes, confirming that acceleration impedes energy extraction away from extremality. Near extremality, Kerr black holes show higher efficiency at lower radii of reconnection locations ($r=1.1-1.14$), while acceleration enhances efficiency across most radii of reconnection locations ($r=1.14-1.24$). This indicates acceleration facilitates energy extraction near extremality. Common to both regimes, accelerating black holes achieve superior efficiency at larger radii of reconnection locations, which is a key advantage. Additionally, the escape condition in Reference \cite{Chen:2024ggq} dictates that at extremely low radii of reconnection locations, $\epsilon_{+}$ cannot reach infinity and instead crosses the event horizon, preventing energy extraction. This particularly affects near-extremal cases ($a=0.998$, $r=1.1-1.14$), in which, while Kerr efficiency dominates, the viable range of radii of reconnection locations is substantially reduced by escape constraints.

\subsection{Permissible Energy Extraction Regions}
\begin{figure}[!h]
  \centering
  \begin{subfigure}{0.45\textwidth}
    \centering
    \includegraphics[width=\linewidth]{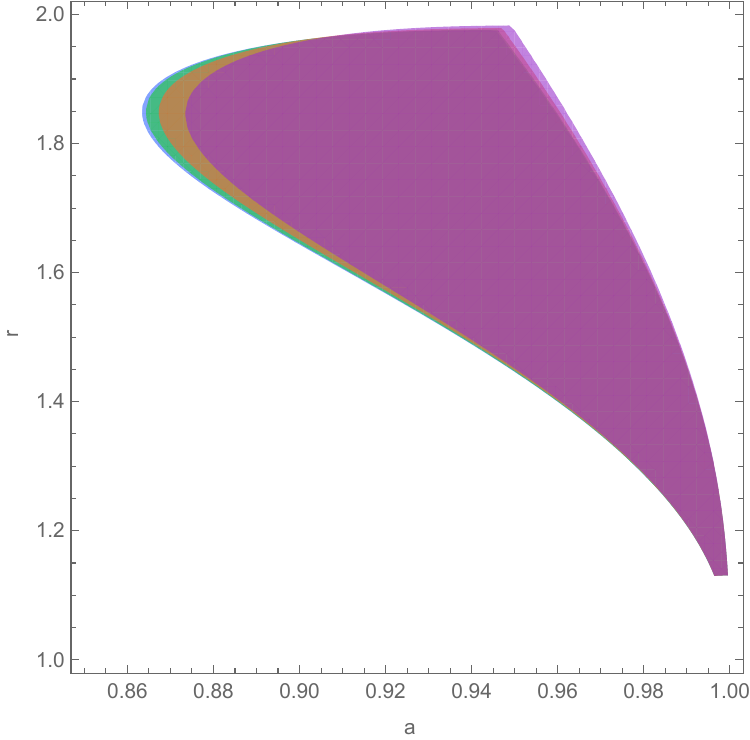}
    \caption{} 
    \label{fig:5a}
  \end{subfigure}
  \begin{subfigure}{0.45\textwidth}
    \centering
    \includegraphics[width=\linewidth]{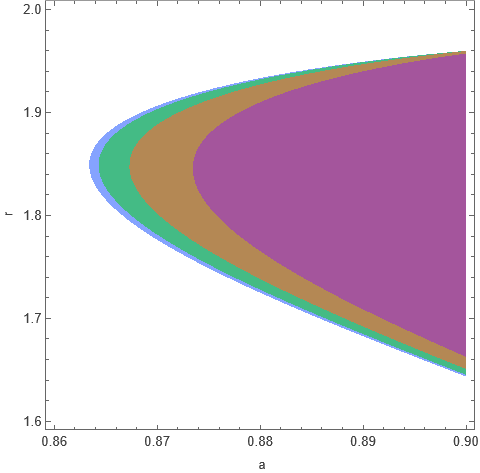} 
    \caption{} 
    \label{fig:5b}
  \end{subfigure}
  \caption{(a) Permissible energy extraction regions in the plunging region. (b) Partial magnification images of (a).}
  \label{fig:5}
\end{figure}

We plot permissible energy extraction regions ($\epsilon_{-} < 0$) in the left panel of Figure \ref{fig:5}, with colors denoting acceleration parameters: blue ($A=0$), green ($A=0.02$), yellow ($A=0.04$), purple ($A=0.06$) at $\sigma=100$. The right panel of Figure \ref{fig:5} shows a magnified view of the left panel. Within the plunging region ($r < r_{\text{ISCO}}$), the escape condition eliminates only minimal areas with the smallest radii of reconnection locations. As established in Section 3.B, acceleration favors higher radii of reconnection locations, making escape constraints negligible\footnote{Consistent with Reference \cite{41}, which omitted escape conditions when analyzing plunging-region azimuthal effects.}.
Figure \ref{fig:5} reveals that the acceleration parameter increases the minimum spin required for energy extraction. Larger acceleration parameters correspond to higher minimum spins, confirming that acceleration impedes energy extraction away from extremality.
\section{ Comisso-Asenjo Process in Circular Orbits of Accelerating Kerr Spacetime}
\subsection{ Comisso-Asenjo Process in Circular Orbits}
For stable circular orbits, $U^{r} = 0$ and $r>$ISCO. According to equation \eqref{23}, $\xi = 0$ follows. In this region, the azimuthal angle serves as a free parameter. Without loss of generality, we set $\xi = 0$, which maximizes both the permissible energy extraction area and efficiency.
The magnetic reconnection formalism in equation \eqref{21} remains applicable. Given $U^{r} = 0$,
\begin{equation}
\hat{v}_{s}^{(r)}=0,\hat{v}_{s}^{(\varphi)}=\hat{v}_{s}.
\end{equation}
Here, equation \eqref{21} becomes identical to equation (28) in Reference \cite{Comisso:2020ykg}. Crucially, our geometric index $g \approx 0.49$ differs from their Sweet-Parker model \cite{63} where $g = 0$.

We now examine $\epsilon_-$ versus $r$ at different $A$ values ($\sigma = 100$) within circular orbits.
\begin{figure}[!h]
  \centering
  \begin{subfigure}{0.45\textwidth}
    \centering
    \includegraphics[width=\linewidth]{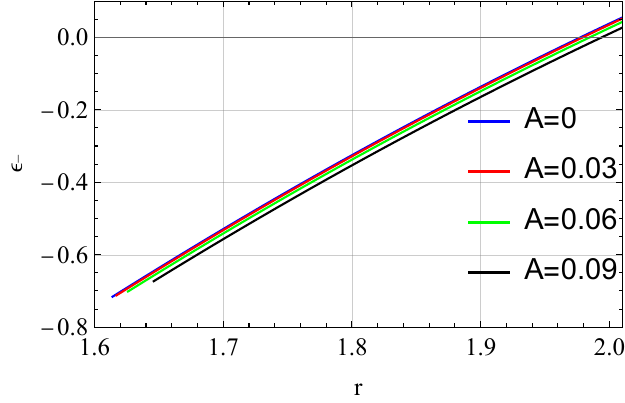}
    \caption{$a=0.98$} 
    \label{fig:6a}
  \end{subfigure}
  \begin{subfigure}{0.45\textwidth}
    \centering
    \includegraphics[width=\linewidth]{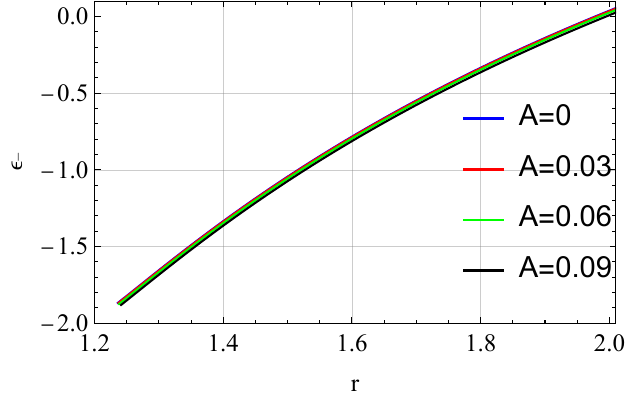} 
    \caption{$a=0.998$} 
    \label{fig:6b}
  \end{subfigure}
  \caption{Variation of $\epsilon_-$ with $r$ for different $A$ values in circular orbits.}
  \label{fig:6}
\end{figure}
Figure \ref{fig:6} demonstrates that in circular orbits, $\epsilon_-$ increases with larger $r$. At fixed $r$, higher acceleration parameters $A$ yield lower $\epsilon_-$ values, indicating more favorable energy extraction. This confirms an extraction advantage for accelerating black holes. The lower $\epsilon_-$ values on the right versus left arise from higher spin parameters on the right.
\subsection{Energy Extraction Efficiency in Circular Orbits}
The efficiency follows equation \eqref{24}. We plot energy extraction efficiency in circular orbits at $\sigma=100$.
\begin{figure}[!h]
  \centering
  \begin{subfigure}{0.45\textwidth}
    \centering
    \includegraphics[width=\linewidth]{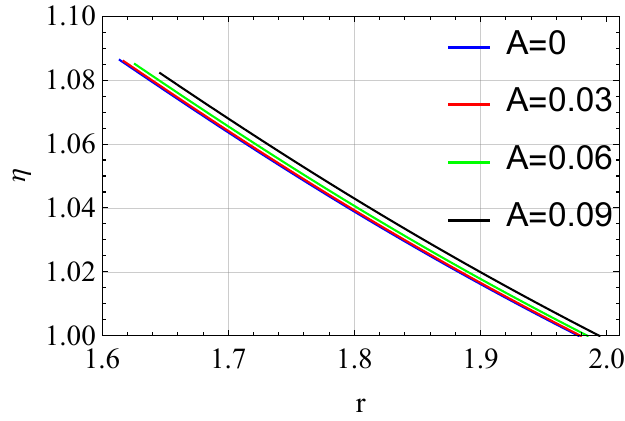}
    \caption{$a=0.98$} 
    \label{fig:7a}
  \end{subfigure}
  \begin{subfigure}{0.45\textwidth}
    \centering
    \includegraphics[width=\linewidth]{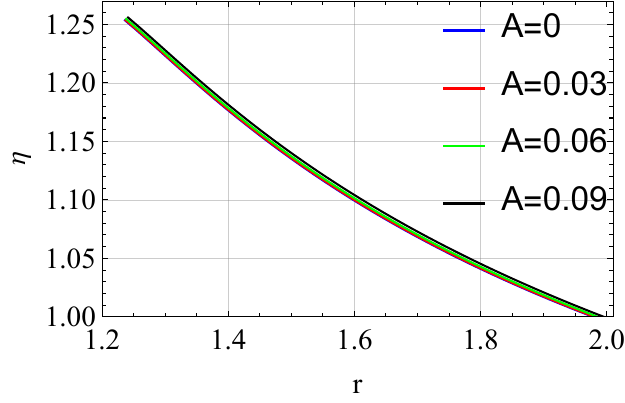} 
    \caption{$a=0.998$} 
    \label{fig:7b}
  \end{subfigure}
  \caption{Variation of energy extraction efficiency with $r$ in circular orbits.}
  \label{fig:7}
\end{figure}
Figure \ref{fig:7} reveals that in circular orbits, $\eta$ decreases with increasing $r$. At fixed $r$, higher acceleration parameters $A$ yield greater $\eta$ values. This demonstrates superior energy extraction capability for accelerating Kerr black holes over standard Kerr black holes in circular orbits, independent of extremality. The higher $\eta$ values on the right versus left arise from elevated spin parameters on the right.

We now compare energy extraction efficiency between circular orbit region   and plunging region. To enable direct comparison, we set the plunging region azimuthal angle equal to the circular orbit value ($\xi=0$) rather than enforcing ideal MHD conditions. Parameters are fixed at $A=0.03$, $\xi=0$, $\sigma=100$, with blue and red curves denoting plunging and circular regions respectively. Additionally, we temporarily relax the requirement $r>$ISCO to include unstable circular orbits at $r<$ISCO. This adjustment is necessary because the plunging region only exists at $r<$ISCO. This treatment applies only in this part, elsewhere circular orbits strictly satisfy $r>$ISCO.
\begin{figure}[!h]
  \centering
  \begin{subfigure}{0.32\textwidth}
    \centering
    \includegraphics[width=\linewidth]{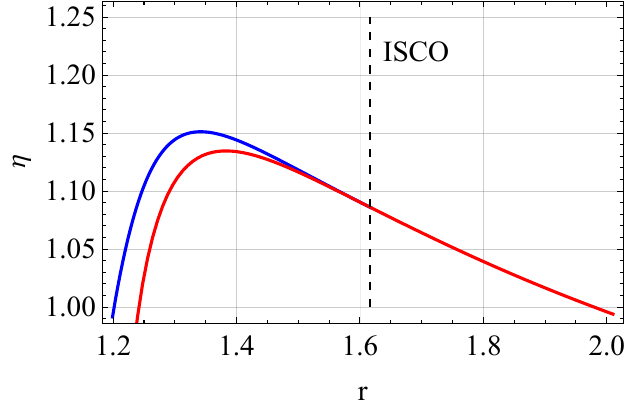}
    \caption{$a=0.98$} 
    \label{fig:9a}
  \end{subfigure}
  \begin{subfigure}{0.32\textwidth}
    \centering
    \includegraphics[width=\linewidth]{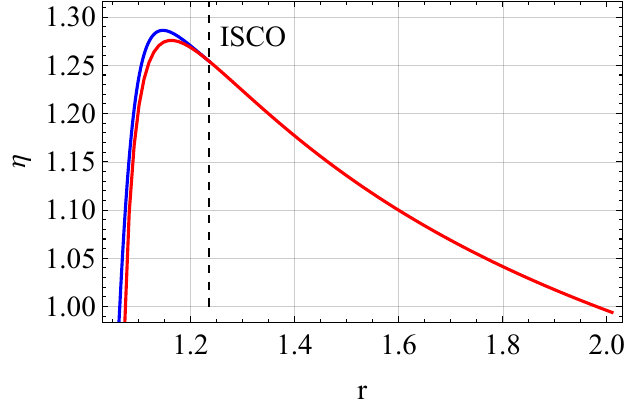} 
    \caption{$a=0.998$} 
    \label{fig:9b}
  \end{subfigure}
  \begin{subfigure}{0.32\textwidth}
    \centering
    \includegraphics[width=\linewidth]{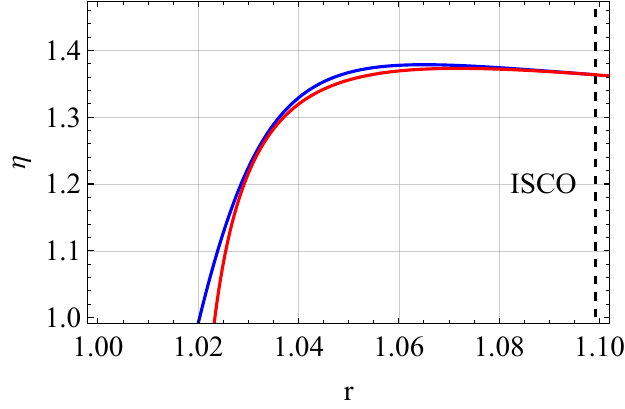} 
    \caption{$a=0.9998$} 
    \label{fig:9c}
  \end{subfigure}
  \caption{Efficiency comparison between plunging and circular orbit regions.}
  \label{fig:9}
\end{figure}

Figure \ref{fig:9} confirms that despite the acceleration parameter, efficiency in the plunging region consistently exceeds that in circular orbits, aligning with established conclusions. It can also be observed that when approaching the extreme black hole, efficiency in the plunging region and the circular orbit region are very similar, almost indistinguishable from each other.

\subsection{Permissible Energy Extraction Regions in Circular Orbits}
\begin{figure}[!h]
  \centering
  \begin{subfigure}{0.32\textwidth}
    \centering
    \includegraphics[width=\linewidth]{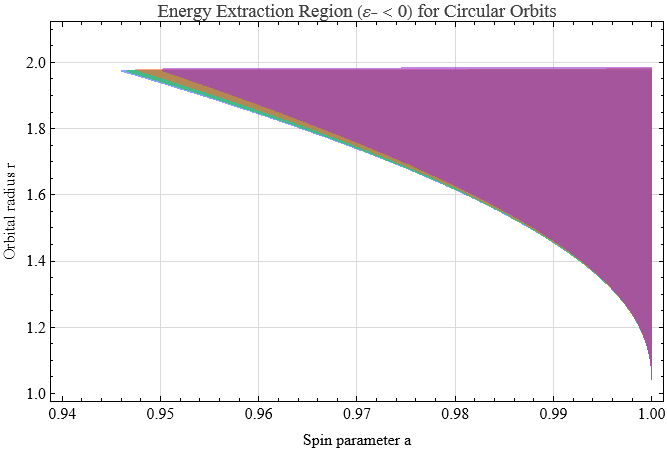}
    \caption{} 
    \label{fig:8a}
  \end{subfigure}
  \begin{subfigure}{0.32\textwidth}
    \centering
    \includegraphics[width=\linewidth]{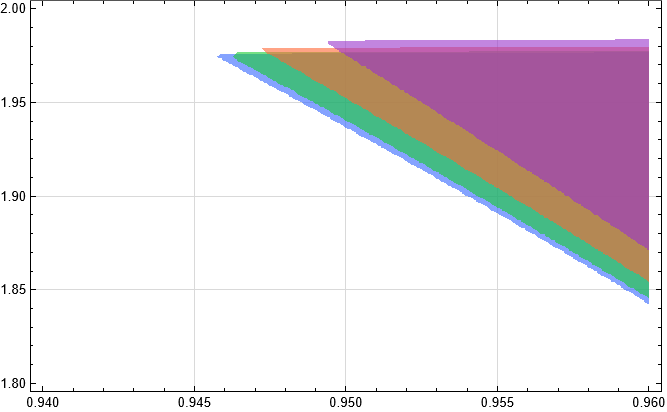} 
    \caption{} 
    \label{fig:8b}
  \end{subfigure}
  \begin{subfigure}{0.32\textwidth}
    \centering
    \includegraphics[width=\linewidth]{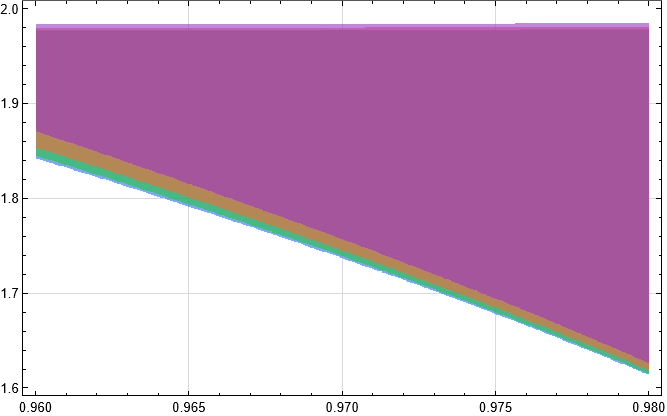} 
    \caption{} 
    \label{fig:8c}
  \end{subfigure}
  \caption{(a) Permissible energy extraction regions in circular orbit region. (b,c) Partial magnification images of (a).}
  \label{fig:8}
\end{figure}

Finally, we plot permissible energy extraction regions for circular orbits in the left panel of Figure \ref{fig:8}, where blue, green, yellow, and purple represent $A=0,0.02,0.04,0.06$ respectively ($\sigma=100$). The middle and right panel of Figure \ref{fig:8} shows a magnified view of the left panel.
Figure \ref{fig:8} demonstrates that as acceleration parameter $A$ increases, the area of permissible energy extraction regions decreases. Nevertheless, energy extraction efficiency in circular orbits surpasses that of Kerr black holes.
\section{Conclusions}
In this work, we investigate energy extraction via the Comisso-Asenjo process in accelerating Kerr black holes. First, we provide a concise overview of accelerating Kerr spacetime, including event horizons, ergospheres, effective potentials, and ISCO. Subsequently, we outline the magnetic reconnection process in the plunging region, where the magnetic field azimuthal angle adheres to the ideal MHD condition, resulting in negative values. Our analysis reveals that compared to Kerr black holes, the acceleration parameter reduces the azimuthal angle magnitude. For viable energy extraction, $\epsilon_-$ must be negative ($\epsilon_- < 0$) and efficiency $\eta$ must exceed unity ($\eta > 1$). We plot $\epsilon_-$ and $\eta$ against $r$ and map permissible extraction regions. For the plunging region, away from extremality the acceleration parameter impedes energy extraction while near extremality it facilitates extraction, particularly at larger radii of reconnection locations. In circular orbits, acceleration reduces permissible extraction areas similarly to the plunging region. Unlike the plunging region however, acceleration enhances extraction efficiency regardless of extremality. Efficiency remains consistently higher in the plunging region than circular orbits, concurring with prior conclusions. The development of the magnetic reconnection Penrose process has been relatively rapid \cite{68,69}. Recent studies have shown that spacetime curvature exerts a significant influence on the properties of magnetic reconnection \cite{64,65,66,67}.
The influence of gravity on the properties of the magnetic reconnection process requires further exploration.

\noindent {\bf Acknowledgments}

\noindent
This work is supported by the National Natural Science Foundation of China (Grants Nos. 12375043,
12575069 ).

\end{document}